\documentclass[11pt]{article}
\usepackage{blois2002,epsfig}
\def\Journal#1#2#3#4{{#1} {\bf #2}, #3 (#4)}

\def\PLB{{\em Phys. Lett.}  B}

\def\PRD{{\em Phys. Rev.} D}

\def\be{\begin{equation}}
\def\ee{\end{equation}}
\def\bea{\begin{eqnarray}}
\def\eea{\end{eqnarray}}

\begin{document}
\vspace*{4cm}
\title{FERMION MASSES AND MIXING IN THICK BRANES}

\author{ M. N. REBELO }

\address{Grupo de F\' \i sica de Part\' \i culas (GFP) - 
Dept. de F\' \i sica,\\
Instituto Superior T\' ecnico, Av. Rovisco Pais, P-1049-001 Lisboa,
Portugal}

\maketitle\abstracts{Fermion masses and mixing, both in the quark
and leptonic sector, are discussed within the approach to the
Yukawa puzzle proposed by Arkani-Hamed and Schmaltz. In the quark
sector we have shown that at least two extra dimensions
are necessary in order to obtain sufficient CP violation,
while reproducing the correct quark mass spectrum
and mixing angles. We have also studied the consequences
of suppressing lepton number violating charged lepton decays
within this scenario for lepton masses and mixing angles. }

\section{Introduction}
This work summarizes the results obtained in 
Branco {\it et al}~\cite{Branco:2000rb} 
and Barenboim {\it et al}~\cite{Barenboim:2001wy}
in the context of a new approach to the flavour puzzle
proposed by Arkani-Hamed and Schmaltz~\cite{Arkani-Hamed:1999dc} (AS)
in the framework of large extra dimensions~\cite{extradims}.

AS suggested that we live in a thick four dimensional subspace
(thick brane) which is infinite in the usual four spacetime
dimensions and possesses a finite volume in the extra 
orthogonal dimensions. Standard Model (SM) fields are
constrained to live on this thick brane whilst gravity
and possibly other gauge singlets propagate in the whole
extradimensional spacetime; furthermore, in this scenario, 
the Higgs boson and the gauge fields are free to propagate
in the entire thick brane, fermions, on the other hand,
have higher dimensional wave functions which are localized
in specific points in the extra dimensions. In this framework
the effective four dimensional Yukawa couplings as well as
all effective couplings involving, for instance, four fermions
are suppressed by exponential factors that depend on the distance
among the different fermion fields localized in the brane. 
This is an important result since in these models there is no 
high energy scale above a few Tev responsible for the
suppression of unobserved phenomena such as proton
decay and lepton number violating charged lepton decays.
Direct coupling between fermions are exponentially suppressed
by the small overlap of their wavefunctions which are given
by narrow Gaussians.

The philosophy behind the AS scenario differs from the traditional
explanation of fermion masses and mixing based on flavour 
symmetries~\cite{flavour} relying instead on the assumption that
the necessary suppression and hierarchy of couplings comes from
the localization of these fiels in the brane. Symmetries are thus 
replaced by geometry with all higher dimensional couplings assumed
do be of order one. It should be noted that the AS scenario
leads to startling consequences~\cite{AGS} which may be observed
at next generation collider experiments.
 
\section{The quark sector}
The effective four-dimensional Yukawa coupling among $Q$, denoting
a quark $SU(2)$ doublet and $U$ an up-type antiquark  $SU(2)$
singlet is given in this framework by 
$\lambda=\kappa~e^{-\mu^2 (l_q-l_u)^2/2}$
with $1/(\sqrt{2}\mu)$ the Gaussian width of the fermionic
fields $Q$ and $U$ whose Gaussian wave functions are centered at 
$l_q$ and $l_u$, respectively and $\kappa$  the higher-dimensional 
Yukawa coupling. Likewise for the coupling among $Q$ and $D$, 
with $D$ denoting the down-type antiquark $SU(2)$ singlet with 
$l_u$ replaced by $l_d$.

Since in the AS framework quark fields are localized in
different places, families are distinguishable, at least
in principle, even in the limit where all Yukawa couplings
vanish. However, one can still refer to different choices
of weak basis (WB) which should be understood as corresponding
to different assumptions about the underlying physics.

In the AS approach it is quite natural to obtain effective
zeros in the Yukawa matrices, since they correspond
to elements which connect fermions ``far'' apart. On the
other hand, equalities or specific relations among elements
of the Yukawa matrices usually require fine-tuning.

The question of whether there is any geometrical configuration 
of quark fields which fits all quark masses and mixing angles
with all $\kappa_{ij}'s$ of order one was addressed by
Mirabelli and Schmaltz~\cite{MS}, and the answer is positive. 
Therefore, without assuming any flavour symmetry it
is possible to accommodate the observed pattern of
fermion masses and mixing angles simply by 
appropriately placing each quark field in a different position.
In Ref~\cite{Branco:2000rb} Branco, Gouv\^ ea and the author 
addressed the issue of CP violation which was not discussed in 
Ref~\cite{MS} 
and showed that, under these assumptions, 
it is not possible,
with only one extra dimension, to obtain sufficient CP violation.
The analysis was done with the help of the WB invariant
condition for CP conservation $Tr[H_u,H_d]^3=0$, derived in 
Ref~ \cite{CPV_cond}, where $H_u\equiv M_uM_u^{\dagger}$ 
and $H_d\equiv M_dM_d^{\dagger}$; this condition is
very useful since it allows
to determine whether or not there is CP violation
without the need of performing the diagonalization
of the mass matrices. 
It was also shown by Chang and Ng~\cite{Chang:2002ww}
that one can have sufficient CP violation with only one
extra dimension at the price of introducing two different Yukawa
coupling strengths for the up-type and down-type quarks.
The search for a solution with two extra dimensions was done
in the nearest neighbour interaction (NNI) basis (exemplified
by the pattern of the mass matrices in Eq.~(\ref{choice})). 
Notice that there is no loss of generality in choosing both  $M_u$ and 
$M_d$ of the NNI form~\cite{NNI}.
An interesting set of locations for the quark
fields leading to the correct spectrum of quark masses
and pattern of mixing angles was found, 
allowing for the right strength of 
CP violation. The locations of the quark fields in the two extra 
dimensions are depicted in Fig.~1 
\begin{figure}
\label{fig:2-dim}
    \centerline{
    \psfig{file=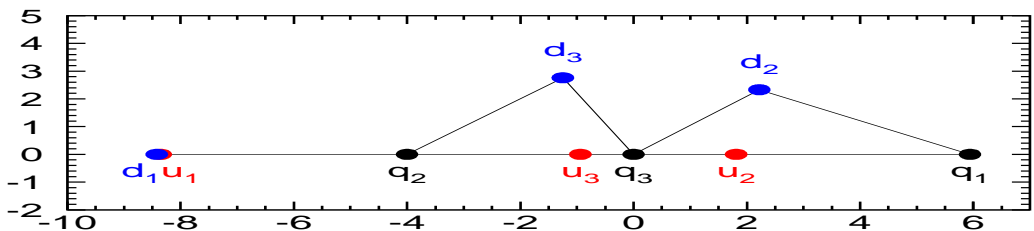,width=1\columnwidth, height=4cm}
}
    \caption{Locations of the quark wave-functions corresponding to
Eq.~(\ref{2d_points}). 
Distances are measured in units of $\mu^{-1}$ (see text).
Once the 
$u_i$'s and $q_i$'s are placed on the same straight line, $d_2$ and $d_3$
are forced into the second dimension.}
\end{figure}
and correspond to:
\begin{equation}
q_i = \frac{1}{\mu} \left( \begin{array}{c}5.941; 0 \\  
-4.008; 0 \\ 
0;0  \end{array} \right), 
u_i = \frac{1}{\mu}  \left( \begin{array}{c}-8.347; 0 \\
1.815; 0 \\
-0.941; 0 \end{array} \right), 
d_i =  \frac{1}{\mu} \left( \begin{array}{c}-8.421; 0 \\
2.219; 2.332 \\
-1.253; 2.767 \end{array} \right),
\label{2d_points}
\end{equation}
assuming $\kappa^{ij} v=1.5m_t$, 
for all $i$ and $j$ and $m_t = 166000$ Mev 
(see Ref~\cite{MS} for details regarding this choice) it  
leads to the mass matrices:
\begin{equation}
M_d=\left( \begin{array}{ccc} 0 & 16.112 & 0 \\
14.690 & 0 & 121.77 \\ 0 & 1400 & 2467.8 \end {array} \right) 
\; \mbox{MeV}, \ \ 
M_u=\left( \begin{array}{ccc} 0 & 50.0 & 0 \\
20.3 & 0 & 2258 \\ 0 & 48 000 & 160 000 \end {array} \right)
\; \mbox{MeV},
\label{choice}
\end{equation}
where the zeros correspond to strongly suppressed matrix elements.

CP violation requires complex entries in the mass terms. In the
NNI basis it is possible to eliminate all complex phases from 
$H_u$, while the off diagonal elements of $H_d$ still have
arbitrary phases, by making a transformation of the type
\begin{equation}
H_u\rightarrow K^{\dagger}H_uK; \hspace{1cm} 
H_d\rightarrow K^{\dagger}H_dK,
\end{equation}
where $K$ is an unitary diagonal matrix. After this transformation
$H_d$ is left with two phases which can be factored out
into another unitary diagonal matrix $K ^\prime$ 
($H_d = K^{\prime \dagger}{H_d}^r K^\prime$ with ${H_d}^r$ real).
One may choose, without loss of generality  
$K'={\rm diag}(1,e^{-i\phi},e^{-i\sigma})$. It was verified that,
in this example, 
the values $\phi = 85^{\circ}$ and $\sigma = 0^{\circ}$
lead to the correct amount of CP violation together with
values for masses and mixing within the experimental range. 
This solution and its rationale is analysed in 
detail in the original reference. It is not possible
in such a short contribution to go into further details. 

\section{The leptonic sector}
In Ref~\cite{Barenboim:2001wy} Barenboim, Branco, Gouv\^ ea and 
the author studied the consequences of suppressing lepton number 
violating charged lepton decays within this scenario for
lepton masses and mixing angles. Due to limitations of space 
we just present here the main conclusions obtained in the reference given
above. 

In the first part of the paper the analysis was done 
in the framework of the SM
with three generations and the addition of three righthanded
singlet neutrino fields which were assumed to be localized
at different points in the extra dimensions. Furthermore lepton number
was also imposed as a conserved symmetry, in order to forbid 
Majorana masses for both the right-handed neutrinos and the
active neutrinos. It was found that the branching ratios 
for flavour violating 
muon and tau decays can be very easily suppressed to levels 
below the current 
experimental bounds, and that very small 
neutrino masses can be obtained for separations of order $10\mu^{-1}$.
Only configurations which yield hierarchical neutrino masses were found.
This is an interesting property of these scenarios. Almost
degenerate neutrino masses have extremely interesting consequences
in terms of mixing and CP violation~\cite{Branco:1998bw} as well as
in the prediction of a bound for neutrinoless double beta 
decay~\cite{Pascoli:2002ae}.
In the AS scenario,
the large neutrino mixing which has been observed in the 
atmospheric neutrino data~\cite{KK} requires fine-tuning of
distances which in principle would be unrelated.  
This is a peculiar feature of the AS scenario for fermion masses: 
it naturally accommodates large mass hierarchies and small mixing 
angles, while it seems to require additional structure 
(since different pairs of fields have to be separated by very similar
distances) in order to explain large mixing angles.
The need for fine-tuning in these solutions should be
interpreted as an additional
challenge in the search for localizing mechanisms. It should
be stressed that the AS scenario is particularly suited for
explaining the absence of lepton flavour violating muon and tau decays,
and can also explain why neutrinos are more than ten orders
of magnitude lighter than the top quark, if right-handed
neutrinos are introduced in the brane, as opposed to scenarios
with bulk neutrinos. 

At the end of the paper
the effect of explaining the absence of flavour changing 
tau and muon decays and charged lepton masses in the AS scenario 
on theories where small
Majorana neutrino masses are generated by breaking lepton number 
in a far away brane was discussed.
In this case, in order to obtain large mixing in the atmospheric 
sector, we were forced
to place different lepton doublet fields so close that
the current
experimental upper bounds on the branching ratios of some rare tau 
and muon decays
were almost saturated~\cite{PDG}. Therefore, these rare decays are expected 
to be observed in the 
next round of experiments, if such a scenario were 
indeed realised in nature. 
Furthermore, very hierarchical neutrino mass-squared differences 
were not attainable, meaning that the solar neutrino puzzle would 
have to be solved either by the SMA or the LMA  solutions.

\section*{Acknowledgments}
The author is grateful to the Theory Division of CERN where
both papers summarized in this contribution were done.
These two works as well as travel and local expenses for
the participation at ``14th Rencontres De Blois''  
received partial support from Funda\c c\~ ao para a Ci\^ encia e a 
Tecnologia (Portugal) through Projects CERN/P/FIS/15 184/1999,
POCTI/36288/FIS/2000 (which includes funds from  EC-FEDER programme),
CERN/FIS/43793/2001, CFIF- Plurianual (2/91) and also from
the European Commission under the RTN contract HPRN-CT-2000-00149.

\end{document}